\documentclass[twocolumn]{aastex62}

\usepackage{graphics,epsf}
\usepackage{amsmath}                % American Mathematical Society package
\usepackage{amsfonts}               % American Mathematical Society fonts
\usepackage{amssymb}                % American Mathematical Society symbol
\usepackage{epsfig}                 % EPS figures
\usepackage{graphicx}
\usepackage{float}
\usepackage{color}
\usepackage[para,online,flushleft]{threeparttable}
\def \s{~\rm{s}}
\def \km{~\rm{km}}

\def \AU{~\rm{AU}}
\def \erg{~\rm{erg}}

\def \yr{~\rm{yr}}

\begin{document}

\title{Possible post-kick jets in SN 1987A} 

%% \correspondingauthor{Noam Soker}
%% \email{soker@physics.technion.ac.il}

%% \author{Efrat Sabach}
%%% \affiliation{Department of Physics, Technion, Haifa, 3200003, Israel}

\author[0000-0003-0375-8987]{Noam Soker}
%\affil{Departmeמt of Physics, Technion, Haifa 3200003, Israel}
\affiliation{Department of Physics, Technion, Haifa, 3200003, Israel; soker@physics.technion.ac.il}
\affiliation{Guangdong Technion Israel Institute of Technology, Shantou 515069, Guangdong Province, China}

%% \author{Rbert T. Fisher}
%% \affiliation{Department of Physics, University of Massachusetts  Dartmouth, 285 Old Westport Road, North Dartmouth, MA 02740, USA; rfisher1@umassd.edu}

\begin{abstract}
I suggest that the recently observed hot-dust elongated structure that likely engulfs the neutron star (NS) remnant of supernova (SN) 1987A was shaped by jets that the NS launched shortly after it acquired its natal kick velocity. I take the axis of the two post-kick jets to be along the long dimension of the hot-dust elongated structure, which I term the bipolar lobe. The jittering jets explosion mechanism accounts for the misalignment of the post-kick jets axis and the main-jets axis of the ejecta. For post-kick jets to shape the bipolar lobe their energy should have been about equal to the energy of the ejecta inner to the outer edge of the lobe,  $E_{\rm 2j} \approx E_{\rm ej}(R_{\rm lobe}) \simeq 4.6 \times 10^{48} \erg$. For an efficiency of ${\zeta}=0.01 -0.1$ to convert accretion energy to post-kick jets' energy I estimate that the post-kick accreted mass was $M_{\rm acc,pk} \approx 2 \times 10^{-4} - 0.002 M_\odot$. The negative jets feedback mechanism, where jets remove some of the mass that could have been accreted, accounts for the limited amount of accreted mass. 
This study adds to the growing recognition of the importance of jets in core collapse supernovae, including post-explosion jets. 
\end{abstract}

\keywords{stars: massive -- stars: neutron -- supernovae: general -- stars: jets}

% ==========================================================
\section{Introduction}
\label{sec:intro}
% ==========================================================

The expansion of the supernova (SN) remnant of SN 1987A (SNR 1987A) allows observations in recent years to better resolve the geometrical structure of the ejecta (e.g., \citealt{Franketal2016, Larssonetal2019a, Micelietal2019, Arendtetal2020}), as well as to establish the properties of dust (e.g., \citealt{DwekArendt2015, Matsuuraetal2019}). 
These observations reveal a complicated non-spherical morphology of the ejecta. 
It is accepted that a binary companion to the progenitor of SN 1987A might account for the non-spherical explosion itself \citep{ChevalierSoker1989}, for the progenitor being a blue supergiant  (e.g., \citealt{Podsiadlowskietal1990, MenonHeger2017, Urushibataetal2018}), and for the formation of the three rings  (e.g., \citealt{Soker1999, MorrisPodsiadlowski2009}).
However, unlike the three circumstellar rings that have a general axisymetrical morphology (e.g., \citealt{Wampleretal1990,Burrowsetal1995}), the structure of the ejecta is much more complicated than just being axi-symmetrical (e.g., \citealt{Abellanetal2017, Matsuuraetal2017}).  
 
A recent addition to the observations that show the non-spherical SNR 1987A inner structure is the detection of a hot-dust blob (\citealt{Ciganetal2019blob}) that might reveal the location of the NS remnant of SN 1987A (\citealt{Ciganetal2019blob, Pageetal2020}), solving the earlier puzzling non-detection of the remnant (e.g., \citealt{Haberletal2006, Indebetouwetal2014}). The hot-dust blob is surrounded by a larger hot-dust region that I term the bipolar lobe. The study of possible implications of this bipolar lobe is the focus of the present paper.  

There are two explosion mechanisms that researchers use to account for the non-spherical structure of SNR 1987A. In the neutrino-driven explosion mechanism instabilities of different kinds alone account for all aspects of the the non-spherical structure (e.g., \citealt{Wongwathanarat2017, Utrobinetal2019, Jerkstrandetal2020}). 
On the other hand, some studies argue that in addition to the role of instabilities a global bipolar outflow (\citealt{Orlandoetal2020}), or several jets that the newly born NS launched at and shortly after the explosion and in varying directions (\citealt{Soker2017a, BearSoker2018SN1987A}) play the major role in determining the large scale departure from spherical symmetry. The jittering jets explosion mechanism (e.g., \citealt{PapishSoker2011, GilkisSoker2014, GilkisSoker2016}) is behind the varying directions of jets' axes. 
Earlier studies attributed some morphological features, e.g., two opposite protrusions (`Eras'), in several SNRs to jittering jets during the explosion (e.g., \citealt{BearSoker2017b, Bearetal2017, GrichenerSoker2017, Akashietal2018}).
The neutrino-driven explosion mechanism and the jittering jets explosion mechanism might overlap, at least in some cases (\citealt{Soker2019JitteringSim, ObergaulingerAloy2020}. 
I note that \cite{Wangetal2002} already suggested that jets powered the explosion of SN 1987A, but their predicted jets' axis is not compatible with new observations by, e.g.,  \cite{Abellanetal2017}. More generally, there are many earlier studies related to CCSN explosions in the frame of the classical jet-driven magnetorotational mechanism, where the  jets are stable and do not jitter (e.g., \citealt{Khokhlovetal1999, Aloyetal2000, Maedaetal2012, LopezCamaraetal2013, BrombergTchekhovskoy2016,  Nishimuraetal2017}). 

The jittering jets explosion mechanism has many challenges to overcome. Regarding the kick velocity of the NS that is relevant to this study, the jittering jets explosion mechanism attributes the kick to the same mechanism as that of the neutrino driven mechanism, the tug-boat
mechanism \citep{Nordhausetal2010, Janka2017}. In this mechanism one or more dense clumps that are expelled by the explosion gravitationally attract the NS and
accelerate it. The jittering jets explosion mechanism predicts that the kick velocity direction tends to avoid the close angles with the main jets axis \citep{BearSoker2018kick}. 

I here raise the possibility that the newly detected hot-dust  bipolar lobe was shaped by post-kick jets, i.e., jets that the NS launched after it acquired its natal kick velocity (section \ref{sec:PostKickJets}). I summarise the main results in section \ref{sec:Summary}. 
I open by discussing the natal kick direction in relation to the main axis of the ejecta (section \ref{sec:KickDirection})

% ==========================================================
\section{Natal kick direction}
\label{sec:KickDirection}
% ==========================================================

\begin{figure*}%[ht!]  % {figure*} to have figure on two columns
%	\centering
% [trim=left bottom right top, . ..]
\includegraphics[trim=0.0cm 5.2cm 0.0cm 2.0cm ,clip, scale=0.8]{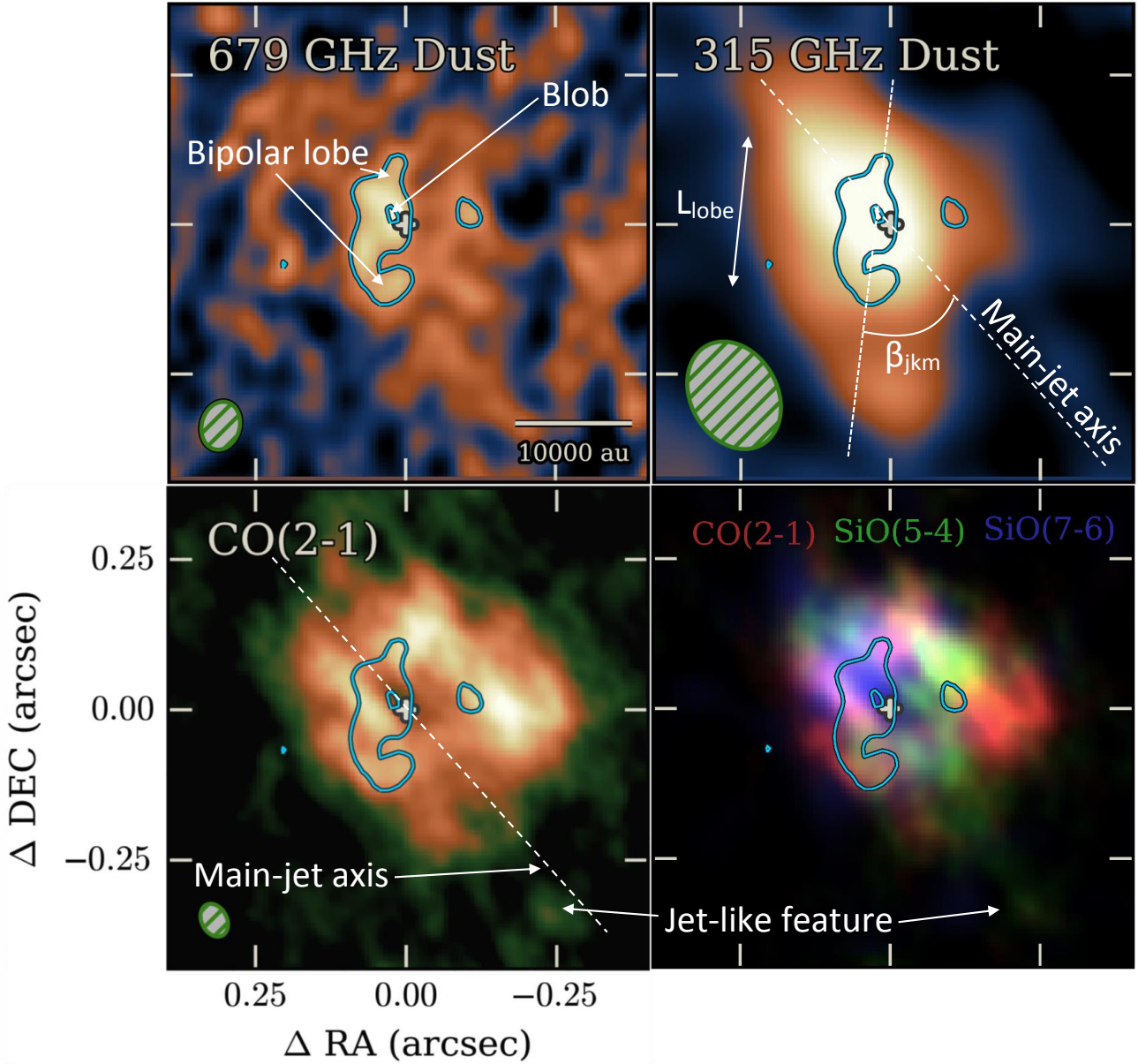}
	\caption{Four panels from figure 3 of \cite{Ciganetal2019blob}, with marking of relevant features that I added in white. The term blob is from \cite{Ciganetal2019blob}, the jet-like feature is according to \cite{BearSoker2018SN1987A}, while the main-jet axis, the bipolar lobe, the length of the bipolar lobe $L_{\rm lobe}$, and $\beta_{\rm jkm}$ are definitions that I introduce here.  
I suggest that post-kick jets shaped the bipolar lobe. 	}
	\label{fig:schematic}
\end{figure*}

In Fig. \ref{fig:schematic} I mark the relevant morphological features for this study, the main-jet axis, the blob, and the bipolar lobe (hereafter the lobe). I use four panels from figure 3 of  \cite{Ciganetal2019blob}. 
\cite{BearSoker2018SN1987A} identify the jet-like feature (lower-right panel of Fig. \ref{fig:schematic}) in the molecular emission image from \cite{Abellanetal2017}. I take the line more or less along this jet-like feature to be the main-jet axis (dashed-white line), as it also defines a general symmetry line of the ejecta. It is very important to keep in mind that in the jittering jets explosion mechanism there is no single jet axis because different jets-launching episodes have different directions.
I define the main-jet axis by eye inspection of the different images of SN 1987A, and in comparison with images of planetary nebulae that show more details and clearer symmetry axes (for comparison of SNRs with planetary nebulae see \citealt{BearSoker2017b, Bearetal2017, Akashietal2018}).  
%FFFFFFFFFFFFFFFFFFFFFFFFFFFFFFFFFFFFFFFFFFFFFFFFFF 
%FFFFFFFFFFFFFFFFFFFFFFFFFFFFFFFFFFFFFFFFFFFFFFFFF 
 
The main-jet axis is more or less in the plane of the sky as the jet-like feature is \citep{Abellanetal2017}.     
If the NS remnant resides at the center of the blob \citep{Ciganetal2019blob, Pageetal2020}, then the component of the natal kick (hereafter kick) velocity on the plane of the sky is more or less along the main-jet axis and to the north (Fig. \ref{fig:schematic}). However, \cite{Pageetal2020} infer the kick direction to be about $30^\circ$ to the observer direction. They base their claim in part on simulations that show the kick velocity to be opposite to the velocity of the iron-group and intermediate-mass elements (e.g., \citealt{Wongwathanaratetal2013}). This is based on simulations of neutrino-driven explosions, but the same holds in the jittering jets explosion mechanism.   
The above kick direction implies that the angle between the main-jet axis and the kick velocity is $\alpha_{\rm jk} \approx 60^\circ$. 
This is compatible with the finding of \cite{BearSoker2018kick} that the angles between the kick velocity and the main-jet axis in SNRs tend to avoid small angles of $\alpha_{\rm jk} \la 40^\circ$. The tendency for large values of $\alpha_{\rm jk}$ is one of the expectations of the jittering jets explosion mechanism \citep{BearSoker2018kick}.
 
\cite{Ciganetal2019blob} and \cite{Pageetal2020} convincingly argue that the NS remnant of SN~1987A heats the dust in the blob. The dust in the much larger structure of the lobe is also hotter than the rest of the ejecta, and so the NS heats also this dust. The lobe has a clumpy structure. 
I will assume below that post-kick jets, namely, jets that the NS launched after it acquired its kick velocity, shaped the lobe to its bipolar morphology. \cite{AkashiSoker2020lobes} present numerical simulations that show how post-explosion jets form a  bipolar structure in the inner part of the ejecta of core collapse supernovae. 
I take the direction of the axis of the two opposite post-kick jets to be more or less along the long dimension of the bipolar lobe.  The jets inflate bubbles, one on each side of the equatorial plane of the post-kick jets, as in many other astrophysical objects, such that the NS remnant radiation can heat the lobe. 

In the jittering jets explosion mechanism the jets' axes of the different jets-launching episodes can be at large angles to each other, but are not completely random to each other. \cite{PapishSoker2014b} argued that the jets' axes of consecutive bipolar jets tend to be in the same plane as the plane of the first two bipolar jet episodes. Still, there is jittering. For the above, it is expected that the axis of the post-kick jets is not along the main-jet axis. The projected angle between them is $\beta_{\rm jkm} \simeq  50^\circ$.
      
% ==========================================================
\section{Post-kick jets}
\label{sec:PostKickJets}
% ==========================================================

% =========================
\subsection{The ejecta}
\label{subsec:Ejecta}
% =========================

Although the ejecta is not spherical, to allow an analytical calculation I take the density profile to be that of a spherical explosion according to \cite{ChevalierSoker1989} as in equations 1-6 of \cite{SuzukiMaeda2019} with $\delta=1$ and $m=10$ 
\begin{equation}
\rho (r, t) = \begin{cases}
        \rho_0 \left( \frac{r}{t v_{\rm br}} \right)^{-1} 
        & r\leq t v_{\rm br}
        \\
        \rho_0 \left( \frac{r}{t v_{\rm br}} \right)^{-10} 
        & r>t v_{\rm br}, 
        \end{cases}
\label{eq:density_profile}
\end{equation}
    \newline
where $M_{\rm ej}$ is the ejecta mass, $E_{\rm SN}$ is its kinetic energy,  
\begin{eqnarray}
\begin{aligned} 
& v_{\rm br} = \left( \frac{20}{7} \right)^{1/2} \left( \frac {E_{\rm SN}}{M_{\rm ej}} \right)^{1/2} 
=3.92 \times 10^3 % 3923.1 
\\& 
\times \left( \frac {E_{\rm SN}}{1.5 \times 10^{51} \erg} \right)^{1/2}
\left( \frac {M_{\rm ej}}{14 M_\odot} \right)^{-1/2}
\km \s^{-1} ,
\end{aligned}
\label{eq:vbr}
\end{eqnarray}
and
\begin{equation}
\rho_0 = \frac {7 M_{\rm ej}}{18 \pi v^3_{\rm br} t^3} .
\label{eq:rho0}
\end{equation}
I scale the ejecta mass and energy with values from \cite{Jerkstrandetal2020}. 

The ALMA observations that \cite{Ciganetal2019blob} present are from 10352 to 10441 days after the explosion of SN~1987A. The ejecta at the break velocity $v_{\rm br}$ is at a distance of $R_{\rm br} (28.5 \yr) = 2.36 \times 10^4 \AU = 0.46 \arcsec$. Namely, it has already crossed the ring as one can see in the H$\alpha$ images (e.g.,  \citealt{Ciganetal2019blob}).    %0.458

% =========================
\subsection{The bipolar lobe}
\label{subsec:Lobe}
% =========================

\cite{Chevalier1989} already studied the possibility that a NS accretes mass from the inner ejecta over months after explosion. He further mentioned the possibility of accretion disk formation. He did not study the role of jets. I here study the role of jets in shaping the recently observed bipolar lobe. 

I assume that jets that the NS remnant of SN~1987A launched after it acquired its kick velocity, i.e., post-kick jets, shaped the lobe to have its elongated structure. Below I crudely estimate the energy of the post-kick jets $E_{\rm 2j,pk}$, and the mass that the NS accreted after it acquired its kick velocity, $M_{\rm acc,pk}$. 

The scenario I propose has the following properties. 
(a) The post-kick jets are also post-explosion jets, in the sense that earlier jets already unbound the star from the newly born NS. (b) These jets would have been launched even without a natal kick. Namely, the accretion disk that launches the jets would have formed even without the natal kick. (c) The kick velocity might change the angular momentum direction of the accreted mass, and therefore also the direction of the jets. (d) The scenario does not refer to the time when the NS launched the post-kick jets. \cite{Chevalier1989} argued that the fall back process might last for months. However, he did not consider the negative jet feedback mechanism, where jets remove mass from the reservoir that feeds the accretion disk (for a review of the jet feedback mechanism see \citealt{Soker2016Rev}). I expect therefore that the NS launched the jets seconds to minutes after explosion, and after it already acquired its natal kick velocity. I encourage numerical simulations to study this process.        
  
The length of the lobe is $L_{\rm lobe} = 1.29 \times 10^4 \AU = 0.25\arcsec$. %  1.286e4
The average distance from the center to the two opposite ends of the lobe is $R_{\rm lobe} \simeq 6.4 \times 10^3 \AU = 0.125\arcsec$.  This corresponds to an ejecta velocity of 
$v_{\rm ej,l} \simeq 1070 \km \s^{-1}$.
The ejecta mass from the center to the radius $R_{\rm lobe}$ is,  for the above values,  
\begin{eqnarray}
\begin{aligned}
& M_{\rm ej} (R_{\rm lobe}) = \int^{R_{\rm lobe}}_0 \rho_0 \left( \frac{r}{R_{\rm br}} \right)^{-1} 4 \pi r^2 dr 
\\ &
= \frac{7}{9} \left( \frac{v_{\rm ej,l}} {v_{\rm br}} \right)^2 M_{\rm ej} \simeq 0.058 M_{\rm ej} \simeq  0.8 M_\odot. 
\end{aligned}
\label{eq:M(lobe)}
\end{eqnarray}

The total kinetic energy of the ejecta inner to $R_{\rm lobe}$ is
\begin{eqnarray}
\begin{aligned}
& E_{\rm ej} (R_{\rm lobe}) = \int^{R_{\rm lobe}}_0 \rho_0 \left( \frac{r}{R_{\rm br}} \right)^{-1} \frac{1}{2} v^2 4 \pi r^2 dr 
\\ &
= \frac{5}{9} E_{\rm SN}  
\left( \frac{v_{\rm ej,l}} {v_{\rm br}} \right)^4  
\simeq  3.1 \times 10^{-3}    % 3.075 
E_{\rm SN} \simeq 4.6 \times 10^{48} \erg,
\end{aligned}
\label{eq:Eej(lobe)}
\end{eqnarray}
where in the last two steps I substituted for the velocities and for the explosion energy. 

I scale the fraction of accretion energy that the jets carry with the canonical fraction of $\zeta \simeq 0.1$. 
Namely, the jets carry about $10 \%$ of the post-kick accreted mass $M_{\rm acc,pk}$ at about the escape velocity from the NS. 
For the jets to shape the lobe, I assume then that the energy in the jets is roughly equal to the ejecta energy inner to $R_{\rm lobe}$ (equation \ref{eq:Eej(lobe)}), i.e., $E_{\rm 2j,pk} \approx E_{\rm ej}(R_{\rm lobe})$. 
The post-kick jets scenario that I propose here requires then that the post-kick mass that the NS remnant of SN~1987A accreted was 
\begin{eqnarray}
\begin{aligned}
M_{\rm acc,pk} \approx
& 1.6 \times 10^{-4}
\left( \frac {E_{\rm 2j} }{5 \times 10^{48} \erg} \right) 
\left( \frac {\zeta}{0.1} \right)^{-1} 
\\ &
\times
\left( \frac {M_{\rm NS}}{1.4 M_\odot} \right)^{-1}
\left( \frac {R_{\rm NS}}{12 \km} \right)  M_\odot . 
\end{aligned}
\label{eq:Maccpk}
\end{eqnarray}

\cite{Chevalier1989} estimated that the central star of SN~1987A could have accreted up to $\approx 0.1 M_\odot$ in the months after explosion. He even mentioned the likely formation of an accretion disk, but he did not include the negative feedback due to the energy that the jets, which the accretion disk is likely to launch, can deposit into the ejecta. I expect that because of the jet feedback mechanism the accreted mass onto the NS remnant of SN~1987A was significantly lower than $0.1 M_\odot$. I actually take the presence of the bipolar lobe in the SNR~1987A to indicate that a negative jet feedback mechanism did take place, like in many other astrophysical systems that have bipolar structures \citep{Soker2016Rev}.  

% ==========================================================
\section{Summary}
\label{sec:Summary}
% ==========================================================

I studied the recently observed hot-dust region in SNR~1987A \citep{Ciganetal2019blob}, which I term the bipolar-lobe (Fig. \ref{fig:schematic}), in the frame of the jittering jets explosion mechanism. I assumed that the radiation from the NS remnant of SN~1987A heats the bipolar lobe (\citealt{Ciganetal2019blob, Pageetal2020}), and therefore that the dust in the lobe has a clear path to radiation from the NS remnant. 

I proposed that post-kick jets that the NS remnant of SN~1987A launched after it acquired its kick velocity shaped the lobe to its bipolar structure. I took the axis of the post-kick jets to be along the long dimension of the bipolar lobe. The jittering jets explosion mechanism accounts for the misalignment of the post-kick jets axis and the main-jets axis (Fig. \ref{fig:schematic}). 

For the jets to shape the bipolar lobe I estimated the energy of the jets to be about equal to the kinetic energy of the ejecta inner to the outer edge of the lobe $R_{\rm lobe} = 0.5 L_{\rm lobe}$. This energy is $E_{\rm 2j} \approx E_{\rm ej}(R_{\rm lobe}) \simeq 4.6 \times 10^{48} \erg$ (eq. \ref{eq:Eej(lobe)}). 
 To launch such jets, the post-kick NS should have accreted a mass
according to equation (\ref{eq:Maccpk}). For an efficiency ${\zeta}=0.01 -0.1$ to convert accretion energy to jets' energy the post-kick accreted mass is crudely $M_{\rm acc,pk} \approx 2 \times 10^{-4} - 0.002 M_\odot$. 

There is an increasing recognition of the importance of post-explosion accretion disks that fall back material forms around the central NS or black hole, including outflows from the accretion disks (e.g., \citealt{AkashiSoker2020lobes, Liuetal2020, Zenatietal2020} for very recent papers). The jittering jets explosion mechanism that attributes the explosion itself to jets, accounts also for post-explosion jets. The main claim of the present study that post-kick jets operated in SN~1987A, leaving the imprint of the bipolar lobe, adds to the growing recognition of the importance of jets in core collapse supernovae. 

% ===================================================
\section*{Acknowledgments}
% ===================================================
 
I thank Avishai Gilkis, Amit Kashi and an anonymous referee for useful comments. This research was supported by a grant from the Israel Science Foundation (420/16 and 769/20) and a grant from the Asher Space Research Fund at the Technion.
% =======================

\end{document}